\documentclass{llncs}
   \usepackage{times}
   \usepackage{mathptm}
   \usepackage{epsfig}
   \usepackage{color}
   \usepackage{latexsym}
   \usepackage{pslatex}
   \usepackage{algorithmic}
   \usepackage{algorithm}

\begin{document}
\pagestyle{headings}  

\title{A SAT-Based Algorithm for Computing Attractors in Synchronous Boolean Networks}
\author{Elena Dubrova and Maxim Teslenko}
\institute{Royal Institute of Technology (KTH), Electrum 229, 164 46 Kista, Sweden\\
\email{ \{dubrova,maximt\}@kth.se}}

\maketitle

\begin{abstract}
This paper addresses the problem of finding cycles in the state transition graphs
of synchronous Boolean networks.  Synchronous Boolean networks are
a class of deterministic finite state machines which are used for the
modeling of gene regulatory networks. Their state transition graph cycles,
called {\em attractors}, represent cell types of the organism being
modeled. When the effect of a disease or a mutation on an organism is
studied, attractors have to be re-computed every time a fault is
injected in the model.  We present an algorithm for finding attractors
which uses a SAT-based bounded model checking. Novel
features of the algorithm compared to the traditional SAT-based bounded model 
checking approaches are: (1) a termination condition
which does not require an explicit computation of the diameter and (2)
a technique to reduce the number of additional clauses which are needed to make
paths loop-free.  
The presented algorithm uses much less space than existing BDD-based approaches
and has a potential to handle several orders of magnitude larger networks.
\end{abstract}


\section{Introduction}

A gene regulatory network (GRN) is a collection of DNA segments in a
cell, called {\em genes}, which interact with each other~\cite{AlBLRRW94}. Each
gene contains information that determine what the gene does and when
the gene is active, or {\em expressed}.  When a gene is active a
process called {\em transcription} takes place, producing an {\em
ribonucleic acid (RNA)} copy of the gene's information. This piece of
RNA can then direct the synthesis of {\em proteins}.  RNA or protein
molecules resulting from the transcription process are known as {\em gene
products}.

Mathematical models of GRNs have been developed to capture the behavior of organisms being modeled.
Common GRN models include ordinary and partial differential equations,
Boolean networks and their generalizations, Petri nets, Bayesian
networks, stochastic equations, and process calculi~\cite{ScB07}.  There is
always tension between generality and level of details, and thus
tractability, of a model.  The most appropriate mathematical framework
can be selected depending on the scale involved, the nature of the
available information, and the problem studied. In this paper, we
consider the Boolean network model, which has been shown useful for
exploring GRNs in the context of cellular
differentiation, cell cycle regulation, 
immune response, and evolution (see~\cite{AlCK03} for an overview).

The {\em Boolean network} is a discrete-space discrete-time model in
which every gene is viewed as a vertex whose input values represent gene
products and output values represent the level of gene expression~\cite{Ka69}.
Th edges between vertices represent the interactions between genes.  These
interactions can be {\em activatory}, with an increase in the
concentration of gene products in one gene leading to an increase in
the level of gene expression in other gene, or {\em inhibitory},
with an increase in one leading to a decrease in the other. 
The nature of influence of regulators on a gene is reflected by the Boolean function assigned to
the vertex. In this paper we consider 
the synchronous type of Boolean networks 
in which the values of functions of all vertices are updated simultaneously
at each time step.

Synchronous Boolean networks can be considered as
a class of deterministic finite state machines.  Any
sequence of consecutive states of a network eventually converges to either a
single state, or a cycle of states, called {\em attractor}.
Attractors represent the pattern of gene expressions in the
corresponding cell types of the organism being modeled~\cite{Ka93}.
When the effect of a disease or a mutation on an organism is
studied, attractors have to be re-computed every time a fault is
injected in the model~\cite{AlCK03}. 

All algorithms for computing attractors in Boolean networks face a
state-space explosion problem that must be addressed to handle
large-scale models. A common approach to combat it is to use symbolic
algorithms which avoid building the state transition graph describing
the dynamic behavior of a GRN. Instead, the state transition graph is
represented implicitly by means of Binary Decision Diagrams (BDDs)~\cite{bryant}.
Algorithms based on BDDs~\cite{DuTM05,GaXMG07,NaTC07} are usually able to process GRN models
with up to a hundred of state variables. However, for larger
networks, BDDs become too memory-consuming.
Simulation-based approaches~\cite{Wu00,BlS01,SoK02}
can be applied to large networks, however, they are incomplete.

Propositional decision procedures (SAT) do not suffer from the potential space explosion
of BDDs and can handle propositional satisﬁability problems with thousands of variables~\cite{DaP60}.
This work is the first step in applying SAT procedures to computing attractors.
The presented approach is based on SAT-based bounded model checking~\cite{BiCCFZ99}.
We use a SAT-solver for identification of paths of a particular length $k$
in the state transition graph of a Boolean network. First we generate a propositional formula representing 
an unfolding of the transition relation of the network 
for $k$ steps. A satisfying assignment to this propositional formula
corresponds to a valid path in the state transition graph.  The process is repeated iteratively 
for larger and larger values of $k$ until all attractors are identified.

Note that systems which we are dealing with are deterministic 
and therefore the problem we are addressing is simpler than the traditional bounded 
model checking. This allows us to use a simple termination condition
which does not require an explicit computation of the diameter and also 
to reduce the number of additional clauses which are needed to make
paths loop-free.  

This paper contributes to the ongoing work on finding attractors 
by providing a complete solution which uses much less space than BDD-based approaches
and thus has a potential to handle several orders of magnitude larger networks.
The existing simulation- and BDD-based algorithms for finding attractors are only applicable to
simple organisms such as the yeast, a flower, or a fruit fly.
The presented approach opens a possibility for exploring much more complex organisms including 
humans.

The paper is organized as follows. Section~\ref{rbn} gives a background on synchronous Boolean networks.
In Section~\ref{ide} we describe the intuitive idea behind the presented approach. Section~\ref{alg}
presents the algorithm. Section~\ref{exp} summarizes experimental results. Section~\ref{con} concludes the
paper and discusses open problems.

\section{The Boolean Network Model} \label{rbn}

In the {\em Boolean network} model of GRNs~\cite{Ka69},
every gene is represented by a vertex $i$ in a directed graph with an
associated state variable $x_i$ that takes the value 1
if the gene is expressed and 0 otherwise. An edge
from one vertex to another indicates that the former gene regulates
the latter. Each vertex $i$ also has an associated Boolean function $f_i$
which reflects the nature of influence
of its regulators.

\begin{figure}[t!]
\begin{center}
\resizebox{0.9\columnwidth}{!} {\input{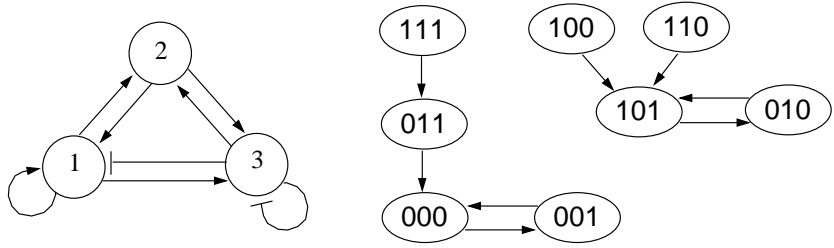}}
\caption{A 3-vertex Boolean network (left) and its state transition graph (right).}\label{f1}
\end{center}
\end{figure}

Time is viewed as proceeding in discrete steps. For the {\em synchronous}
type of update, the expressions of all genes are changed simultaneously.
At each time step,  the next value of the variable $x_i$ 
 is determined by current values of $r$ regulators of $i$ as:
\begin{equation} \label{eq1}
x_i^{+}= f_i(x_{i_1}, x_{i_2}, \ldots, x_{i_{r_i}}).
\end{equation}
where $x_{i_1},x_{i_2},\ldots,x_{i_{r_i}}$ are state variables associated to the regulators of
$x_i$.

The {\em state of the network} is defined as an ordered $n$-tuple of
values of state variables $x_1, x_2, ..., x_n$ 
at a particular moment of time. 
Since a synchronous Boolean network is
deterministic and finite, any sequence of its consecutive
states eventually converges to either a single state, or a cycle of
states, called {\em attractor}.  

An example of a Boolean network with 3 vertices is shown on the left-hand side of Figure~\ref{f1}. Arrows
indicate activatory regulation and blunt-ends represent inhibitory 
regulation\footnote{The use of two types of edges is common for describing Boolean networks in spite of the fact that it is redundant since the type of regulation is fully defined by the associated functions.}
The following Boolean functions are associated to vertices:
\[
\begin{array}{l}
f_1 = \neg x_3 \wedge  (x_1 \vee x_2)  \\
f_2 = x_1 \wedge x_3 \\
f_3 = \neg x_3 \vee  (x_1 \wedge x_2)
\end{array}
\]
where ``$\wedge$'', ``$\vee$'', and ``$\neg$'' stand for the Boolean AND, OR
and NOT operations, respectively.  
The State Transition Graph (STG) of this network 
is shown in the right-hand side of Figure~\ref{f1}. 
It has two attractors of length two.

The Boolean network model has been extended to a multiple-valued one, in which each
variable can take $m$ values rather than only two~\cite{Dub06,GaMXG07a}. Since each
$m$-valued variable can be encoded by $\lceil \log_2 m \rceil$ Boolean
variables, any multiple-valued network can be translated to a
Boolean one and treated using the presented approach with no conceptual
difference. 

\section{Intuitive Idea}   \label{ide}

The presented algorithm searches for a path of a given length $k$ in the STG of a Boolean network.
If a path is found, we check whether it contains a loop or not.
Since each state of the STG of a Boolean network
has a unique next state, once a path reaches a loop, 
it never leaves it. Therefore, we can determine the presence of a loop
simply by checking whether the last state of the path occurs
at least twice.

Clearly, all states between any two occurrences of the last state belong to 
a loop. A loop corresponds to an attractor. We mark all attractor's states.
In the following iterations, 
we will only search for paths in which the last state is not marked.

Until at least one attractor remains unmarked, we can find a path of any length
since we can cycle in an attractor forever.
However, once all attractors are identified and marked, we will only be able find 
paths which are shorter than a given length (at most the diameter of the STG). 
So, when we search for a path of some length $k$ and it does not exist,
this means that all attractors are already identified and the algorithm can terminate.

If a path of length $k$ does exist and it is loop-free, we double $k$ and continue
the search for a path of the new length. 

We illustrate the algorithm on the example of the Boolean network in Figure~\ref{f1}.
The algorithm starts from searching for a path of length $k= 3$. Suppose that the path we found
is $111 \rightarrow 011 \rightarrow 000$. Since the last state $(000)$ occurs only once,
this path is loop-free. We increase $k$ to 6 and continue the search for a path of length 6.
 Suppose that the path we found
is $110 \rightarrow 101 \rightarrow 010 \rightarrow 101 \rightarrow 010 \rightarrow 101$.
Now, we can see that $(101,010)$ is a two-state attractor and mark it. The following
search for a path of length 6 may return us the path $011 \rightarrow 000 \rightarrow 001 \rightarrow 000 \rightarrow 001 \rightarrow 000$.
Again, we mark $(000,001)$ is a two-state attractor. The next search shows that
there exist no more paths of length 6. We conclude that all attractors are identified and
terminate the algorithm. 

As another possibility, while searching for a path of length $k= 3$ we may find a path
$001 \rightarrow 000 \rightarrow 001$. In this case, we mark $(000,001)$ is a two-state attractor
and continue the search for a path of length 3. Next we may find the path $010 \rightarrow 101 \rightarrow 010$, and mark it.
The following search will show that there exist no more paths of length 3 and algorithm will terminate.

As we can see, the presented algorithm may terminate either 
before or after the depth of
unfolding becomes equal to the diameter of the STG.

\section{Description of the Algorithm}   \label{alg}

The pseudocode of the presented algorithm is given as
Algorithm~\ref{alg1}.  We use a SAT-solver for identification of paths of a particular length $k$
in the STG of a Boolean network. First we generate a propositional formula $F$ representing 
an unfolding of the transition relation $T$ of the network 
for $k$ steps. A satisfying assignment to this propositional formula
corresponds to a valid path in the STG.  The process is repeated iteratively 
for larger and larger values of $k$ until all attractors are identified.

\subsection{Initial unfolding}

Given a Boolean network with $n$ vertices and the
transition relation $T$, the algorithm first unfolds the transition
relation $k$ times, where $k = min(n, 100)$. 
We empirically found it more time-efficient to unfold 
the transition relation directly by $n$ steps for small networks of size $n < 100$.
For large networks with $n > 1000$, unfolding by $n$ steps might take too much memory and
it is usually unnecessary for identification of all attractors. This is justified by
some specific features of gene regulatory networks which we describe
in Section~\ref{exp}.

\subsection{Direction of unfolding}

In the pseudocode, we use $T_{p \ldots r}$ to denote the
transition relation $T$ which is unfolded from the time step $p$ to the time
step $r$, i.e.
\[
T_{p \ldots r} = \bigwedge_{i=p}^{r-1}  T(s_i,s_{i+1}).
\]
where $s_i$ denotes the state of a Boolean network at the time step $i$.

\begin{algorithm}[t]
\caption{An algorithm for computing attractors in a Boolean network with $n$ vertices and the transition relation $T$.} \label{alg1}
\begin{algorithmic}
\STATE $i = n$
\STATE attractor\_is\_found = False
\STATE $A(s_0) = 0$ ~~~~ /* $A(s_0)$ is the set of states of all attractors expressed in terms of variables of $s_0$ */
\STATE $F = T_{-i \ldots 0}$ ~~ /* $F$ is the propositional formula representing the unfolding $T_{-i \ldots 0}$ */
\WHILE {Sat(F)}    
\FOR {$(j = -1; j \geq -i; j--)$}
\IF {$s_j = s_0$}
\FOR {$(k = 0; k > j; k--)$}
\STATE $A(s_0) = A(s_0) \cup (s_0 \leftrightarrow c_k)$  /* $c_k \in \{0,1\}^n$ is an assignment of the variables */ 
\STATE ~~~~~~~~~~~~~~~~~~~~ /* of $s_k$ returned by the SAT-solver; $s_0 \leftrightarrow c_k$ is defined by (\ref{sk}) */
\ENDFOR
\STATE $F = F \wedge \neg A(s_0)$
\STATE break
\ENDIF
\ENDFOR
\IF {attractor\_is\_found}
\STATE attractor\_is\_found = False
\ELSE
\STATE $F = F \wedge T_{-2*i \ldots -i}$
\STATE $i = 2*i$
\ENDIF
\ENDWHILE
\end{algorithmic}
\end{algorithm}

One specific feature of our algorithm is that we always unfold $T$ from some time step $-p$ 
to the time step $0$ so that 
the previous time frames
rather then the next ones are added to the unfolding. The depth of the unfolding is increased by
decreasing $-p$. In this way, the last state of the unfolded transition relation is always 
$s_0$, independently of the depth of the unfolding. Later we will explain how 
this helps us to reduce the number of additional clauses which are needed to make
paths loop-free.  

\subsection{Identification of paths}

Once the transition relation is unfolded, a SAT-solver is called to find a satisfying assignment for the
resulting propositional formula $F$. The function Sat in the
pseudocode corresponds to a call to a SAT-solver. Sat takes an
expression and returns True if there exists an assignment of variables
which make the whole expression true. 

If a satisfying assignment does not exist, this means that there is
no path of length $i$ in the STG. This
implies that all attractors have been already identified and marked
in the STG, so the algorithm terminates.

\subsection{Checking paths for loops}

If a SAT-solver finds a satisfying assignment, the algorithm checks whether 
there is a loop in the path corresponding to this assignment.
As we mentioned before, we can determine the presence of a loop
by checking whether the last state of a path occurs
at least twice. Since in our case the last state of any unfolded
transition relation is $s_0$, to identify an attractor, it is
sufficient to check weather $s_0$ occurs at least twice on a path.

\subsection{Adding restrictions to $F$}

If $s_j = s_0$ for some $j \in \{-i, \ldots, -2, -1\}$, then we can conclude that we found an
attractor of length $j$. In this case, each of the
attractor's states is added to the a characteristic function $A(s_0)$
which represents the set of states of all attractors expressed in
terms of variables of $s_0$.

The $n$-bit vector $c_i \in \{0,1\}^n$ is used to denote an assignment of variables of $s_k$ which is returned by 
the SAT-solver. The notation $s_0 \leftrightarrow c_k$ means that 
\begin{equation} \label{sk}
s_0 \leftrightarrow c_k =  \bigwedge_{i=0}^{n-1} (s_0[i]  \leftrightarrow c_k[i]),
\end{equation}
where $s_0[i]$ is $i$th variable of the state $s_0$ and $c_k[i]$ is $i$th bit-position of the vector $c_i$.

By adding $\neg A(s_0)$ to the propositional formula $F$ we constrain
$F$ in such a way that any satisfying assignment for $F$ will contain no states of
already identified attractors.  Note that, in our case, it is sufficient to ensure that the
state $s_0$ does not belong to any already identified attractor in order to
guaranty that no state in this path belongs to an identified attractor.
Now it becomes evident why we have chosen to make the 
last state of any unfolded transition relation $s_0$.

\begin{table}[t!]\centering
\begin{tabular}{|c|c|c ||c|c||c|c|c|} \hline
Benchmark      & vertices	 &  Attractors  &\multicolumn{2}{c|}{BDD-based~\cite{DuTM05}} & \multicolumn{3}{c|}{SAT-based}  \\ \cline{4-8}
name     		&  $n$	& number $\times$ length & sec  & MB &  sec  & MB & unfolding depth  \\ \hline
Arabidopsis thaliana & 15 & $10 \times 1$ & 0.077 & 19.14 & 0.035 & 1.76 & 15 \\
Budding yeast & 12 & $7 \times 1$ & 0.109 & 19.82 & 0.046 & 1.91 & 24 \\ 
Drosophila melanogaster & 52 &  $7 \times 1$& - & $>$ 1000 & 0.093 & 2.32 & 52 \\
Fission yeast & 10 & $13 \times 1$ & 0.062 & 19.04 & 0.030 & 1.78 & 10 \\
Mammalian cell & 10 &  $1 \times 1, 1 \times 7$ & 0.060 & 19.04 & 0.028 & 1.76 & 10 \\
T-cell receptor & 40 & $8 \times 1, 1 \times 6$ & 0.093 & 19.34 & 0.030 & 1.98 & 40 \\
T-helper cell & 23 &  $3 \times 1$ & 0.107 & 19.61 & 0.042 & 1.81 & 23 \\ \hline
\end{tabular}
\caption{Experimental results for the Boolean networks models of real organisms; "-" stands for memory blow up.} \label{t1}
\end{table}

\section{Experimental Results}   \label{exp}

We have implemented an experimental tool based on the presented algorithm. 
In this section, we compare it to the BDD-based tool for finding attractors\footnote{Both tools are available at http://web.it.kth.se/$\sim$dubrova.} from~\cite{DuTM05}.
Our implementation uses MiniSAT SAT-solver~\cite{EeS05}. 
All experiments were run on a PC with Pentium III
750 MHz processor and 256 Mb memory. 

As benchmarks\footnote{At present there is no common set of benchmarks for the GRN simulation tools. We have constructed the input descriptions of models shown in Table~\ref{t1} manually from the data in the corresponding papers.} we use existing Boolean networks models of real 
organisms shown in Table~\ref{t1}: 
control of flower morphogenesis in {\em Arabidopsis thaliana}~\cite{chaos2006}, 
budding yeast cell cycle regulation~\cite{citeulike:1048296}, 
{\em Drosophila melanogaster} segment polarity genes expression patterns prediction~\cite{AlO03}, 
fission yeast cell cycle regulation~\cite{Davidich_Bornholdt_BooleanNetworkCellCycle_2008}, 
the mammalian cell cycle regulation~\cite{FaNCT06}, 
T-cell receptor signaling pathway analysis~\cite{citeulike:562504},  and
T-helper cell differentiation~\cite{citeulike:558097}. 

As we can see from Table~\ref{t1}, the presented SAT-based algorithm
uses an order of magnitude less space than the BDD-based algorithm. For the 
{\em Drosophila melanogaster}, the BDD-based algorithm runs out of memory.
The memory blow up occurs while trying to construct the initial transition relation $T$.

We can also see that for only one benchmark, budding yeast, the depth of unfolding
had to be doubled to $2n$. For the rest of benchmarks, all attractors were identified
after the first unfolding by $n$ steps.

The performance of the presented algorithm is determined by the number and
length of attractors in a network, as well as to the length of the longest path to
an attractor.
For large networks, we may expect  the number of attractors to be considerably smaller than 
the number of vertices $n$ of the network. This is because the number of 
vertices $n$ in a Boolean network corresponds to the number of relevant genes in  
the organism it models, and the number of attractors $N_a$ corresponds to the number of cell types
of this organisms. 
Different hypothesizes have been made suggesting that, for large Boolean networks, 
$N_a = O(\sqrt n)$~\cite{Ka93} or $N_a = O(n^{2/3})$~\cite{BlS01}.

As we can see from Table~\ref{t1}, the length of attractors is usually one. This is because
the states of attractors represent  the expression levels of genes in a given cell type,
which are normally stable~\cite{AlO03}. So, we may expect the length of attractors to be
a small constant for large networks as well.

Finally, the longest path to an attractor is related to the time which takes a cell to settle down 
into a stable pattern in the process of cell differentiation~\cite{Ka93}. 
For all benchmarks in Table~\ref{t1} this parameter was smaller than $n$.
No empirical results are known for larger networks.

\section{Conclusion}   \label{con}

This work is the first step in applying SAT procedures to finding attractors in Boolean networks.
We believe that the presented approach has a potential to handle several orders of magnitude 
larger networks than the ones which can be handled by the BDD-based approaches.
Unfortunately, existing Boolean models of real organisms are small, so they do not allow us to
support this claim.
Our next step is to work in collaboration with biologists on applying the 
presented tool to create larger models of more complex organisms.

\bibliographystyle{ieeetr}
\bibliography{bib}

\end{document}